\documentclass[twocolumn,showpacs,preprintnumbers,amsmath,amssymb,floatfix]{revtex4}
\usepackage{amsmath}
\usepackage{graphicx}
\def\ra{\rangle}
\def\la{\langle}
\def\C{\rm C}

\begin{document}
\bibliographystyle{apsrev}



\title{Splitting Sensitivity of the Ground  and 7.6 eV Isomeric States  of $^{229}$Th}
\author{A.\ C.\ Hayes, J.\ L.\ Friar and P. M\"oller}
\affiliation{Theoretical Division, Los Alamos National Laboratory, 
Los Alamos, New Mexico 87545}
\date{\today}

\begin{abstract}

The lowest-known excited state in nuclei is the 7.6 eV isomer of
$^{229}$Th. This energy is within the range of laser-based investigations that
could allow accurate measurements of possible temporal variation of this energy
splitting. This in turn could probe temporal variation of the fine-structure
constant or other parameters in the nuclear Hamiltonian. We investigate the
sensitivity of this transition energy to these quantities. We find that the two
states are predicted to have identical deformations and thus the same
Coulomb energies within the accuracy of the model (viz., within roughly 30 keV).  We therefore find no enhanced sensitivity to variation of the fine-structure constant. In the case of the
strong interaction the energy splitting is found to have a complicated
dependence on several parameters of the interaction, which makes an accurate
prediction of sensitivity to temporal changes of fundamental constants
problematical.  Neither the strong- nor Coulomb-interaction contributions to the
energy splitting of this doublet can be constrained within an accuracy better
than a few tens of keV, so that only upper limits can be set on the possible
sensitivity to temporal variations of the fundamental constants.

\end{abstract}
\maketitle

The isotope $^{229}$Th has recently become of interest because of its unusually
low-lying (7.6 eV, 3/2$^+$) \cite{beck} first excited state, which is an isomeric state
with an estimated half-life of 5 hours. The  $5/2^+ - 3/2^+$ ground-state-to-isomer transition
energy is within the range of atomic transitions, and it has been suggested that
this almost degenerate doublet in $^{229}$Th could be used as a nuclear clock
\cite{clock} and as a sensitive probe of possible temporal variation of
fundamental constants, including the fine-structure constant ($\alpha$) and the quark
mass\cite{flam}. The sensitivity of the transition energy to temporal changes in
the fundamental constants varies considerably depending on the assumptions
made\cite{flam,hayes,variation1}. 
For example, in refs. \cite{flam,variation1} temporal variation of the fine 
structure constant ($\dot{\alpha}$) was related to constants in the Nilsson Hamiltonian, whereas
in \cite{hayes} $\dot{\alpha}$ was shown to be proportional to the Coulomb energy 
difference between the two states,
which requires detailed information about the deformations involved.
Thus it is important to understand the
nuclear-structure issues giving rise to this doublet. 

In the present work we
examined  this doublet using the finite-range 
microscopic-macroscopic model (FRDM)\cite{moller-nix}, which describes many
nuclear-structure properties (such as ground-state masses and deformations) over
a broad range of nuclei. Our goal is to examine the sensitivity of the energy
splitting between these states to the underlying components of the effective
nuclear interaction, including the single-particle potential and the pairing, 
spin-orbit, and Coulomb interactions. Knowledge of this
sensitivity is essential for determining the sensitivity of the transition
energy to possible temporal variation of fundamental constants.

In our macroscopic-microscopic model\cite{moller-nix} the macroscopic
terms give  the smooth variation of the nuclear potential energy (mass)
with proton number $Z$, neutron number $N$, and deformation. 
The dependence of nuclear structure properties on  microscopic quantum-mechanical effects is obtained from a deformed single-particle potential through the use of Strutinsky's method\cite{strutinsky67:a,strutinsky68:a}.  Nine constants of the macroscopic part
have been determined in a least-squares adjustment to 1654 measured nuclear
masses with $Z\ge 8$ and $N\ge 8$; the details are given in \cite{moller-nix}.
For these 1654 nuclei ranging from $^{16}$O to $^{263}$106 the model mass
accuracy is 0.669 MeV. For the heavier mass regions the model is more accurate: for N$\ge$65 the corresponding accuracy is 0.448 MeV\@. Values of
other model constants (such as the depth and diffuseness of the single-particle
potential, and the strengths of the spin-orbit and pairing interactions) are
determined from other global considerations, as discussed in \cite{moller-nix}.
Ground-state masses and shapes have been calculated for 8979 nuclei and
tabulated in
\cite{moller-nix}. The shape parameters tabulated are quadrupole ($\epsilon_2$),
octupole ($\epsilon_3$), hexadecapole ($\epsilon_4$), and hexacontatetrapole
($\epsilon_6$) deformation (shape) degrees of freedom.  Strong evidence of
reliability of the model is given by its now well-established predictive capabilities
for new nuclear-mass regions, including unstable nuclei and super-heavy nuclei.

In the present calculations we  carried out a  high-accuracy 
determination of the ground-state deformation and quasi-particle energy by 
 minimizing the potential energy for $^{229}$Th on a fine
deformation grid.  Taking advantage of  enhanced present-day computational
power, we  varied all four
shape parameters 
in steps of 0.001, which is to be compared to the 0.05 
grid step used in \cite{moller-nix}.
We predict three very closely lying states with identical deformation:  the
$5/2^+$ ground state, the $3/2^+$ isomer, and a $5/2^-$  state.
The three lowest (almost degenerate) neutron quasi-particle states are predicted to have deformations:
$\epsilon_2 =0.170$, 
$\epsilon_3 = 0.000$, 
$\epsilon_4=-0.084$, and
$\epsilon_6 =-0.002$.
The $5/2^-$ state, with asymptotic quantum numbers 
$[N n_z\Lambda, K^\pi] = [732, 5/2^-]$, lies lowest in energy.
The ground-state and isomer doublet of interest have asymptotic quantum numbers
 $[6 \, 3 \, 3, 5/2^+]$ and  $[6 \, 3 \, 1,3/2^+]$, respectively, and are predicted to lie at 13.6 keV and 21.6 keV above the $5/2^-$.  

Since the three lowest neutron quasi-particle configurations are predicted to 
have the same deformation (which determines their proton distribution), they
also have the same Coulomb energy (within the model uncertainties). This is not the case for other states in $^{229}$Th, for which the energy was found to be minimized by different values of the shape parameters.

Although minor changes to the model parameters could move this triplet of states relative to one another, the exact position of the 5/2$^-$ state is not essential to our present interest, since the predicted energy of that state does not directly affect the sensitivity of the splitting between the 5/2$^+ -3/2^+$ doublet to the underlying interaction.
Given the global accuracy of the model in this mass region, we expect that 
the 5/2$^-$ state should lie somewhere within 50 keV of the ground state.

The calculated positions of the three lowest-lying
states as  functions of deformation are shown in Fig.~1, and the three states are seen to track very 
closely in energy for values of  $\epsilon_2$ between roughly 0.12 and 0.22. 
As discussed below, we find a similar close tracking for 
all of the deformation parameters:  $\epsilon_2,\cdots,\epsilon_6$. 
For comparison, we also show in Fig.~2 the
$[6 \, 3 \, 1, 3/2^+]$  and  $[6 \, 3 \, 3, 5/2^+]$ 
doublet together with two other
low-lying states of $^{229}$Th. 
Note the very different energy scales used in Figs.~1 and 2, and  that
the latter two states do not show a
minimum at the same deformation as the three lowest states, which is the more typical situation seen 
in deformed nuclei. The energies of different
quasi-particle configurations for a nucleus are not generally minimized by the  same values of deformation
parameters.
 
The  predicted energy separation of the $[6 \, 3 \, 1, 3/2^+]$  and  $[6 \, 3 \, 3, 5/2^+]$ doublet (viz., 8.3 keV)  is large compared to the actual excitation energy of the isomer (7.6 eV),
but it nonetheless means that on a ``normal'' nuclear-energy scale the two
states are predicted to be almost degenerate. For comparison, typical
quasi-single-particle energy splittings in this mass region are a few hundred
keV. For completeness we note that the odd-neutron wave function for the $\Omega^{\pi} =3/2^+$ isomer is calculated to have the following asymptotic Nilsson components:

\begin{eqnarray}  
 |3/2^+\ra = 0.650[633] +0.519[642] +0.431[613]\nonumber \\
-0.221[622]  -0.137[602] -0.131[853] \, ,
\end{eqnarray}
while that of the  $\Omega^{\pi} =5/2^+$ ground state has components:
\begin{eqnarray}  
|5/2^+\ra= 0.656[631] +0.521[642] +0.403[611] \nonumber \\
-0.242[651] -0.123[871] -0.104[622] \, .
 \end{eqnarray}
Those components in the wave functions whose squared amplitudes are less than 1\% are not listed.

When quasi-particle configurations have the same deformation,
the energy difference  between them is given by 
\begin{eqnarray}
\Delta E_{\rm{g.s.}-\rm{iso}} = \sqrt{(\epsilon_{\rm{iso}} -\lambda)^2 
+ \Delta^2} - \sqrt{(\epsilon_{\rm{g.s}.} - \lambda)^2 + \Delta^2}\, , 
\end{eqnarray}
where $\lambda$ is the Fermi energy, $\Delta$ is the pairing gap obtained by
solving the pairing equations, and $\epsilon^i_\nu$ are 
the single-particle energies for the two states; all of these energies depend on deformation. The near
degeneracy between the ground state and isomer of $^{229}$Th arises because the
Fermi surface energy is about midway between $\epsilon_{\rm iso}$ and
$\epsilon_{\rm g.s.}$ . The isomer and the ground state will be exactly degenerate
when $(\epsilon_{\rm{iso}} -\lambda) = -(\epsilon_{\rm{g.s.}} - \lambda)$ (i.e., for $\lambda = (\epsilon_{\rm{iso}} +
\epsilon_{\rm{g.s.}})/2$).

The single-particle energies $\epsilon_\nu$ depend on the deformed
single-particle  potential well and on the spin-orbit interaction.
These in turn depend on parameters
governing the depth and diffuseness of the potential well and the strength
parameter of the spin-orbit interaction.  Although the single-particle
energies depend on the shape of the potential, the shape parameters are not
in the category of adjustable parameters, and single-particle energies are calculated by minimizing the
energy of the chosen configuration.  

The dependence of the doublet splitting on the fine-structure constant is
particularly straightforward to examine in our model, as we discuss below.  On the other hand, expressing that
splitting in terms of more fundamental parameters of the strong interaction (such
as the quark mass, for example) would require a very detailed and non-trivial analysis of the relation between the model parameters and fundamental sub-nucleon degrees of freedom\cite{bob}.  Our analysis is therefore necessarily restricted to a study of the
sensitivity of the predicted doublet splitting to the effective interactions in
our model.


\begin{figure}[t] 
 \begin{center} 
\includegraphics[width=\linewidth]{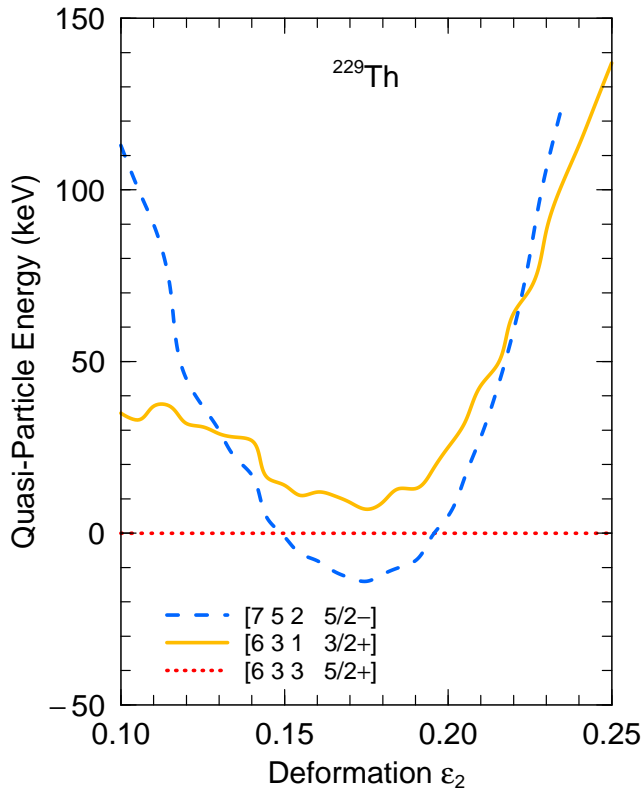} 
\caption{[Color online] Calculated quasi-particle energies for the three lowest quasiparticle
states (relative to  the $[6\,3\,3,5/2^+]$ ground state, which therefore has zero energy for all deformations) in 
$^{229}$Th as functions of $\epsilon_2$. All three states remain almost degenerate at all values of $\epsilon_2$ near the common 
minimum at $\epsilon_2 =0.170.$}
\label{qp3}
 \end{center}
\end{figure}

\begin{figure}[t] 
 \begin{center} 
 \includegraphics[width=\linewidth]{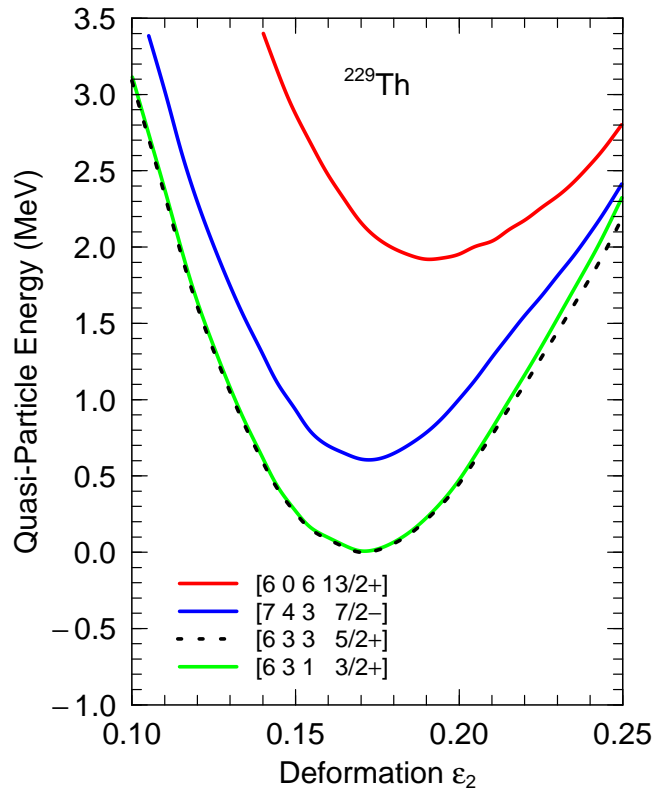} 
\caption{[Color online] Calculated  quasi-particle energies of the 3/2$^+$-5/2$^+$ doublet and of a
$7/2^-$ state  and a $13/2^+$ state in
$^{229}$Th as functions of $\epsilon_2$. The doublet states remain
almost degenerate for a wide range of values of $\epsilon_2$. Note the change in energy scale from Fig~1.}
\label{qp4}
 \end{center}
\end{figure}

The similarity in the shape of the ground and isomeric states of
$^{229}$Th implies that (within the accuracy of the model)  they have the same charge (i.e., proton) distribution and therefore the same Coulomb energy. This has unfortunate implications with respect to experimental searches for a temporal variation in the 
fine-structure constant ($\dot{\alpha}$) obtained from measuring a temporal variation in the energy
splitting between these two states (denoted by $\dot{\omega}$).  The latter variation is proportional to the
Coulomb-energy difference of the two states \cite{hayes}:
\begin{equation}
\dot{\omega} =  \la \la V_{\C} \ra \ra \:  \frac{\dot{\alpha}}{\alpha} \, ,
\end{equation}
where $\la \la V_{\C} \ra \ra = \la \rm{iso}\! \mid V_{\C} \mid \! \rm{iso}\ra - \la
\rm{g.s.}\! \mid V_{\C}\mid \! \rm{g.s.} \ra$.  We therefore find essentially no sensitivity to
$\dot{\alpha}$, since $\la \la V_{\C} \ra \ra = 0$ 
for our chosen mesh-parameter step size. Allowing $\epsilon_2$ to vary by the mesh-parameter step size (0.001)
in this calculation produces a variation in $\la \la V_{\C} \ra \ra $ of approximately 30  keV, and this sets the upper limit for variations in 
$\dot{\omega}$ relative to  $\dot{\alpha}/\alpha$. Uncertainties in the effective nuclear interaction prevent any nuclear-structure calculation from being predictive on an eV scale.

\begin{table*}[t]\label{vary}
\begin{tabular}{| l| l| l| l| l|}
\hline
\multicolumn{5}{|c|}{$\Delta E_{\rm{g.s.}-\rm{iso}}$} \\\hline
 Best & $V_0(+10\%,-10\%)$ & $a(+10\%,-10\%)$ & $\lambda_n(+10\%,-10\%)$ & $G(+10\%,-10\%)$\\
&&&& \\\hline
8.8 keV & -4.4keV, 31.6 keV & 25.4 keV, 2.0 keV & -44.2 keV, 29 keV & 8.7 keV, 9.2 keV\\
&&&& \\\hline
\end{tabular}
\caption{Variation in the predicted 3/2$^+$-5/2$^+$ doublet energy splitting for $^{229}$Th 
induced by a $\pm$10\%
variation of four of the global parameters of the model (see Ref.\cite{moller-nix}). The column labeled
``Best'' is the predicted splitting with the standard values of these
parameters,  $V_0$ is the depth of the single-particle potential, $a$ is the
range or diffuseness of the single-particle potential, $\lambda_n$ is the
strength of the neutron spin-orbit interaction, and $G$ is the pairing
strength. Splittings that are listed with a negative sign mean that 
the  $[6\,3\,3, 5/2^+]$  and $[6\,3\,1, 3/2^+]$ states were inverted.}
\end{table*}

With the exception of accidental degeneracies, nearly degenerate
doublets in deformed nuclei reflect a strong similarity in the deformation of
the states involved. If the deformations of two states are not similar their
energy splittings are typically {\bf  at least} several tens of keV. To illustrate this and
to investigate the accuracy of the model for very close-lying (eV) doublets we
examined the ground and isomeric states in $^{235}$U, where the observed energy
splitting is 76 eV.  We calculated this ($7/2^-, 1/2^+$) doublet, again allowing
all of the deformation parameters to vary independently for each state. The
results are very similar to those for $^{229}$Th, with the ground-state and
isomer energies being minimized by the  same values of the deformation parameters, and lying
just below and above the Fermi surface, respectively. The deformations
for these two states in $^{235}$U were found to be $\epsilon_2=0.205$,
$\epsilon_3=0$, $\epsilon_4=-0.07$ and $\epsilon_6 =0.025$, and the predicted
energy splitting was 29 keV\@.

We next considered the sensitivity of the structure of the $^{229}$Th doublet to
several of the globally determined strong-interaction parameters of the model,
namely, the depth and diffuseness of the single-particle potential, and the
strength of the pairing and spin-orbit interactions. We varied each of these four global
parameters  by $\pm10\%$. We note, however, that variations of this magnitude
would likely destroy the agreement with ground-state masses and for some nuclei would lead to incorrect assignments. 
 For example, the strengths of the neutron and proton spin-orbit
interactions have been fitted to experimental levels in the rare-earth and
actinide regions, where properties of these nuclei  (particularly level
spacings and orderings) strongly suggest a linear dependence of the spin-orbit
strength on the nuclear mass number $A$:
\begin{equation}
\lambda_{\rm n,p} =k_{\rm n,p}\,  A + l_{\rm n,p}\, .
\end{equation}
With this parameterization the constraints on $ k_{\rm n}, k_{\rm p}, l_{\rm n}, l_{\rm p}$ are
considerably tighter than 10\%. Nonetheless, in order to gain a better understanding
of the origin of the near degeneracy in $^{229}$Th we varied the strength of
this and the other three parameters by $\pm$10\%. The results are summarized in
Table 1. We find that the predicted doublet splitting varies on the keV scale
within a 10\% variation of these four global parameters, while for some variations the levels are predicted to be inverted (indicted by a minus sign in Table I). There is no one parameter or
part of the interaction determining or dominating the magnitude of the predicted splitting.
The degeneracy instead arises from a complicated combination of different effects causing the ground state and  isomer to lie just below and above the Fermi surface, respectively.

In summary, we have examined the structure of the ground state and isomer in
$^{229}$Th. Within the present model the two states are predicted to be
different quasi-single-particle neutron states, but corresponding to the same
nuclear deformation.
The Coulomb energy is predicted to be the same for the two states (within an uncertainty of roughly 30 keV induced by our mesh-step size), which
suggests no enhanced sensitivity to $\dot{\alpha}$. The origin of the near
degeneracy appears to be accidental and is difficult to parameterize in terms of
any one component of the effective interaction.
Nevertheless, this doublet remains of interest in possible searches for
time variations in fundamental physics because of the small energy splitting involved.
 
\section{Acknowledgments}
We wish to thank  Naftali Auerbach, John Becker, Victor Flambaum, Steve Lamoreaux, Jerry Wilhelmy, Bob Wiringa, and Xinxin Zhao for useful comments and/or discussions.

\end{document}